\documentclass{aa}
\usepackage{graphicx,psfig}
\usepackage{natbib}
\usepackage{txfonts}
\def\bp{\object{$\beta$\,Pictoris}}

\def\be{\begin{equation}}
\def\ee{\end{equation}}
\begin{document}
\title{Outer edges of debris discs}
\subtitle{how sharp is sharp?}

\author{P. Th\'ebault\inst{1,2}, Y. Wu\inst{3}}
\institute{
Stockholm Observatory, Albanova Universitetcentrum, SE-10691 Stockholm,
Sweden
\and
Observatoire de Paris, Section de Meudon,
F-92195 Meudon Principal Cedex, France
\and 
Department of Astronomy and Astrophysics, University of Toronto,
50 St.George Street, Toronto, ON M5S 3H4, Canada}

\offprints{P. Th\'ebault} \mail{philippe.thebault@obspm.fr}
\date{Received; accepted} \titlerunning{Outer Edges of debris discs}
\authorrunning{Th\'ebault and Wu}

\abstract
%
{
Rings or annulus-like features have been observed in most imaged
debris discs. Outside the main ring, while some systems (e.g., $\beta$ Pictoris and AU Mic)
exhibit smooth surface brightness profiles (SB) that fall off roughly as
$\sim r^{-3.5}$, others (e.g. HR 4796A and HD
139664) display large drops in luminosity at the ring's outer edge and
steeper radial luminosity profiles.
}
%
{
We seek to understand this diversity of outer edge profiles
 under the ``natural'' collisional evolution of the system,
without invoking external agents such as planets or gas.
}
%
{
We use a multi-annulus statistical code to follow the evolution
of a collisional population, ranging in size from dust grains to planetesimals
and initially confined within a belt (the "birth ring"). 
The crucial effect of radiation pressure on the dynamics and
spatial distribution of the smallest grains is taken into account.
We explore the dependence of the resulting
disc surface brightness profile on various parameters.
}
%
{
The disc typically evolves toward a ``standard'' steady state,
where the radial surface brightness profile smoothly 
decreases with radius as $r^{-3.5}$ outside the birth ring. 
This confirms and extends the
semi-analytical study of Strubbe \& Chiang (2006) and provides a firm basis
for interpreting observed discs.
Deviations from this typical profile, in the form of a sharp outer
edge and a steeper fall-off, occur for two "extreme" cases:
1) When the birth ring is so massive that it becomes
radially optically thick for the smallest
grains. However, the required disc mass is probably too high here to be realistic. 
2) When the dynamical excitation of the dust-producing planetesimals is so
low ($<e>$ and $<i> \leq 0.01$) that the smallest grains, which otherwise
dominate the optical depth of the system, are preferentially
depleted. This low-excitation case, although possibly not generic,
cannot be ruled out by observations for most systems, .
}
%
{
Our ``standard'' profile provides a satisfactory explanation for
a large group of debris discs that show smooth outer edges and
$SB \propto r^{-3.5}$. Systems with sharper outer edges, barring other
confining agents, could still be explained by ``natural'' collisional evolution
if their dynamical excitation is very low. We show that such a dynamically-cold
case provides a satisfactory fit to the specific HR4796A ring.
}
\maketitle

\section{Introduction}

\subsection{the ubiquity of ring-like features}

Dusty debris discs have been detected by their infrared excess around
$\sim 15\%$ of nearby main sequence stars \citep[e.g.][]{back93}.
More than a dozen of these discs have also been imaged, mainly in
scattered light, since the initial observation of the $\beta$ Pictoris
system by \citet{smi84}.
One unexpected result from these images is that almost no system displays
a smooth extended radial profile: the usual morphology is the presence
of rings (or annuli) where the bulk of the dust population 
is located. This ring morphology is in fact so common that
\citet{stru06} pointed out that the debris disc phenomenon could more
appropriately be renamed debris ``ring'' phenomenon.  Even for the
archetypal debris ``disc'' $\beta$ Pictoris, which has been imaged
from 5 to a few thousand AU, the bulk of the dust is probably
concentrated in a rather narrow region between 80 and 120 AU
\citep[e.g.][]{aug01}. 
One of the few systems actually resembling an extended ``smooth''
disc might be Vega, for which Spitzer mid--infrared images show a
rather smooth radial luminosity profile
\citep{su05}. At the other end of disc morphologies, among the
most striking ring features are the ones around HR4796
\citep[e.g][]{Jay98,koer98,schnei99}, HD139664 \citep{kal06},
and Fomalhaut \citep{kal05}.

\begin{table*}
\begin{minipage}{\textwidth}
\caption[]{Geometry and surface brightness profile for a selection of
debris disc systems resolved in scattered light images. 
}
\label{init}
\renewcommand{\footnoterule}{}
\begin{tabular}{lcccc}
System \footnote{For many of these systems, many additional features, i.e.,
warps, clumps, etc., have been observed (it is especially
true for \bp) but we focus here on the main issue of average
radial profiles}& Orientation & Detected Radial Extent\footnote{The radial
extents and surface
brightness profiles are given for regions beyond the likely ``birth
ring'' -- the region where scattering luminosity peaks and where most
parent bodies are believed to reside (see Sec.\,\ref{sec:approach})} & Surface Brightness $SB
\propto r^{\alpha}$ & Reference\\
\hline
HD 53143 & $\sim$ face-on & $55-110$AU &
$\alpha \sim -3$ & Kalas et al. (2006)\\
$\beta$ Pic & edge-on & $127-193$AU &
$\alpha \simeq -3$ 
& Golimovski et al.(2006)
\\
& & $193-258$AU&
$\alpha \simeq -4$ & \\
HD 32297 & edge-on & $80-170$AU &
$\alpha \sim -3.5$ & \citet{schnei99}\\
 & & $170-350$AU &
$\alpha \sim -3$ (averaged over the 2 ansae)& \\
AU Mic & edge-on & $32-210$AU & $\alpha \sim -3.8$ (averaged over the 2 ansae) 
& Fitzgerald et al (2007)  \\
HD139664 & edge-on & $83-109$AU&
$\alpha \sim -4.5$ 
& Kalas et. al. (2006)\\
HD 107146 & $\sim$ face-on & $130-185$AU & 
$\alpha \sim -4.8 \pm 0.3$  &
Ardila et al. (2004)\\
HD 181327 & face on & $100-200$ AU &
$\alpha \sim -5$ 
& \citet{schnei06}\\
Fomalhaut &  $66^{o}$ inclination  &  $140-160$ AU 
& $\alpha \sim -6.1$ &
Kalas et al. (2006)\\
HR 4796A & $73^{o}$ inclination & $70-120$AU & 
$\alpha \sim -7.5$
& \citet{wahhaj05} \\
\hline
\end{tabular}
\label{table:observed}
\end{minipage}
\end{table*}
 
The characteristics that most differentiates one ring-like system
from the other is the sharpness of the luminosity drop at the inner and outer
edges of the rings. We shall in this paper focus on the outer edge
issue, for which systems can be basically divided into two
categories \citep[see also][]{kal06}: \begin{itemize}
\item
``Sharp edge'' rings, displaying abrupt surface brightness drops,
sometimes as steep as $r^{-8}$ beyond the ring. The most representative
members of this group are HR4796A and Fomalhaut.
\item
``Smooth edge'' rings, with no sharp outer edge and a surface brightness
drop beyond the ring in $r^{-3}$ or $r^{-4}$. The most famous examples
are here $\beta$ Pic and AU Mic
\end{itemize}
See Tab.\ref{table:observed} for a list of outer-edge profiles
for several resolved debris discs.

The presence of ring features is commonly attributed to some sculpting
mechanisms including, in particular, the presence of a massive planet
(see, for instance, \citet{quillen07} for Fomalhaut,
\citet{frey07} for $\beta$ Pictoris, \citet{wyatt99} for HR4796A,
or the more general study of \citet{moro05}). 
The gravitational effect of such a planet can truncate or create gaps
in the disc, either directly on the dust particles or indirectly on
the planetesimals that produce these particles.

\subsection{outer edges}

Nevertheless, while planets may be a natural and easy explanation for sculpting the
inner edges, outer edges are a different problem. They are difficult to
explain  in the light of one well established fact about debris discs,
i.e. that the observed dust is not primordial but steadily produced
by collisions, through a collisional cascade starting at much
larger parent bodies, maybe in the planetesimal size range \citep[e.g.][]{lag00}.
In this respect, even if there is a sharp outer edge for the parent body
population, collisions would constantly produce very small grains
that would be launched by radiation
pressure onto eccentric or even unbound orbits, thus populating the
region beyond the outer edge and erasing the appearance of a narrow
ring over timescales which might be shorter than those for
gravitational sculpting by a planet.
This issue is a critical one,
since these small grains dominate the total geometric cross section
and thus the flux in scattered light \citep[see the discussion
in][]{theb07}. 

One possible confining mechanism for the sharp outer edge is gas. For
discs transiting between gas-dominated phase (proto-planetary discs)
to dust-dominated phase (debris discs), narrow dust rings may arise
\citep{klahr05}. \citet{beslawu} further demonstrate 
that there exists an instability with which the residual gas collects
grains of various sizes (even those subject to radiation pressure)
into a narrow belt. However, this mechanism requires at least
comparable amount of gas and dust. This may be difficult to justify
for most debris discs, which are evolved systems where the amount of
gas is probably too low to prevent the smallest grains to be launched
onto very eccentric orbits smoothing out any sharp outer edge.

``Razor sharp'' outer edges are thus very difficult to explain
in the presence of this unavoidable outward launching of small grains.
However, although no perfect abrupt outer edge has indeed been observed
\footnote{HD141569A could be a possible exception, but this
system is  likely not a true debris disc, being
significantly younger and gas rich \citep{jonk06}, but 
a member of the loosely defined "transition object" category.}
(unlike inner edges, which are in some cases, like Fomalhaut, almost
razor sharp), a great variety of outer edge profiles
does exist, from relatively smooth to very steep 
(see Tab.\,\ref{table:observed}).

In this study, we address the issue
of how these different profiles can be physically
achieved: can this diversity be explained by the sole ``natural''
evolution of a collisional active disc steadily producing
small, radiation-pressure affected grains, or is
(are) additional mechanism(s) needed?
We consider initial conditions which are {\it a priori}
the most favourable for creating sharp edges, by assuming a
population of large parent bodies confined within an annulus
with an abrupt cutoff at its outer edge.
How this initial confinement may have come about is itself an
interesting question but is not the focus of the current
paper (see however the discussion in Sec.\,\ref{sec:real}).
The outcome we consider as a reference for our investigation
is the radial surface brightness (SB) profile in
scattered light, since this is an observable which is reasonably
well constrained for most imaged debris discs
(either directly observed or obtained by de-projection).
We consider the nominal case of a disc seen edge-on,
but results can easily be extrapolated to face-on systems, since
SB profiles for both orientations tend towards the same radial
dependence far from the birth ring (given the same radial dust
distribution).

In several previous studies, all addressing the specific $\beta$ Pic case
\citep{lec96, aug01, theb05}, it has been
argued that the ``natural'' SB profile outside the collisionaly
active parent bodies belt, or ``birth ring'', falls off as $\sim
r^{-5}$. This is based on the assumption that all particles produced
in the birth ring have a size distribution which scales
as $dN/ds \propto s^{-3.5}$, \citep[as expected for an idealized
infinite collisional cascade at equilibrium, see][]{dohn69}, 
down to the radiation blow--out limit $s=s_{0.5}$ (where
the ratio of radiation pressure to gravity $\beta = 0.5$). 
The smallest radiation pressure-affected grains,
which dominate the light receiving area, are then diluted along their
eccentric orbit. This geometrical spread 
results in $SB \propto r^{-5}$.
However, \citet{stru06} argued that, since high-$\beta$ grains
spend a long time in the collisionaly inactive region beyond the birth
ring, the $dN/ds \propto s^{-3.5}$ collisional equilibrium
law should only apply to the small fraction $f$ of these grains which
are present in the collisionaly active birth ring. This results in
a $1/f$ excess of the disc-integrated number of small grains,
which in turn results in a flatter $SB$ profile in $r^{-3.5}$.
They applied their theory to the AU Mic disc \footnote{AU Mic
is an M star with weak radiation but where stellar wind is believed to
act on small grains in an equivalent way as radiation pressure does
around more massive A stars \citep[e.g][]{aug06}.} and reproduced the
observed SB profile, spectral energy distribution and disc colour.
\citet{stru06} further argued that the observed SB profile depends 
only weakly on the radial and size distributions of grains within the
birth ring. The discs which exhibit a fall-off sharper than $r^{-3.5}$
are thus puzzling in the face of this theory.

The innovative model of \citet{stru06} 
is build on analytical derivations and Monte-Carlo modeling
which did not actually treat the collisional evolution of the system and
relies on several simplifying assumptions.
The main one is that the size distribution is fixed and is
assumed to follow the idealized $dN/ds \propto s^{-3.5}$ scaling
(corrected by the fraction $1/f$), whereas several studies
have shown that this law cannot hold in real systems
\citep[see][and references therein]{theb07}
\footnote{The need for realistic size distributions departing
from fixed power laws has been very recently emphasized by
\citet{fitz07} in their analysis of the AUMic data}.
Another issue is that when evaluating collisional life-times,
only the vertical velocity of the grains was taken into
account, thus neglecting their radial movement which
can be appreciable, if not dominant for the smallest grains.
Finally, the specific dynamics of the small radiation-pressure-affected
grains, in particular the fact that they suffer much more frequent collisions
and at much higher velocities, is not taken into account.

\section{Our Approach}
\label{sec:approach}

We re-address these issues using a numerical approach quantitatively
following the collisional evolution of the full system.
We start with a
birth ring of parent bodies in a perfectly confined annulus and let it
collisionally evolve. The temporal as well as spatial evolution of the
size distribution are followed, taking into account the radial
excursions of high--$\beta$ particles. 
As previously mentioned, we derive for each simulation
the surface brightness profile in scattered light.

\subsection{numerical model} \label{sec:model}

We use a statistical
particle--in--a--box model to follow the evolution in size and spatial
distribution of a population of collisionally interacting bodies. This
code has initially been developed, in its single--annulus version, for the
study of the inner $\beta$ Pic disc
\citep{theb03}, and later upgraded to a multi--annulus version (i.e.,
with 1-D radial resolution) for the study of collisional processes in
extended debris discs \citep{theb07}. The detailed description of the
code can be found in these two papers, and here we recall some of its
main characteristics.

The entire system is spatially divided into $n_a$ concentric radial
annuli. Within each annulus of index $ia$, the solid body population is divided
by size into $n_s$ bins that cover
a broad size range spanning from kilometre to micron. With a standard
$log(2)$ size increment, this requires $n_s \sim100$. Evolution of the
particle number within one $(ia,i)$ bin ($i$ being the size
distribution index) is contributed by all destructive impacts between $(ia,i)$
objects and
bodies from other $(ia',i')$ bins as well as all impacts between other
size bins producing new $(ia,i)$ objects. Collision rates are
estimated statistically. These rates, as well as collision outcomes,
depend on average values of the encounter velocities $\Delta
v_{(ia,i,ia',i')}$.  For objects not subject to significant
radiation pressure, relative velocities are obtained from the
eccentricity and inclination distributions through the classical
expression, valid for randomized orbits
\citep[e.g.][]{lisste93}:
\begin{equation}
\langle \Delta v\rangle_{(ia,i,,i')} =
\left( \frac{5}{4} \langle e^{2}\rangle + \langle
i^{2}\rangle \right)^{1/2}\,\langle v_{kep(ia)}\rangle
\label{dv}
\end{equation}
where $v_{kep(ia)}$ is the Keplerian velocity at radial
distance $r_{ia}$.

For the smallest particles, the effect of radiation pressure,
which places objects on highly eccentric
orbits, is taken into account. Inter annuli interactions,
induced by the significant radial excursion of these bodies are considered,
and $\Delta v_{(ia,i,ia',i')}$ are derived through separate deterministic
N--body runs.

Collision outcomes are divided into 2 types, cratering
and fragmentation, depending on the ratio
between the specific impacting kinetic energy $E_{col}$ and the
specific shaterring energy $Q*$, which depends on object sizes and
composition.
In both regimes, the size distributions of the newly produced
fragments are derived through detailed energy scaling prescriptions,
which are presented at length in \citet{theb03} and \citet{theb07}.
Possible reaccumulation onto the impacting objects is also accounted
for.

\subsection{Set Up}
\label{subsec:parameters}

Our numerical model requires the following inputs: the ring's average
distance from the star, $r_{BR}$, its radial width $\Delta r_{BR}$,
the average free eccentricities $\langle e \rangle$ and inclinations
$\langle i \rangle$ of the
parent bodies (non affected by radiation pressure)
and the initial particles' size distribution and total mass.
We chose to parameterize the latter by $M_{dust}$, the
total mass of objects with sizes $s<s_{d}=1\,$cm, even though our
simulations include bodies up to $10\,$ km, because
the total ``dust'' mass is a parameter which can often
be constrained from observations. 
We also assume equipartition between eccentricity and inclination,
so that $\langle e \rangle = 2 \langle i \rangle$, at least for
the parent bodies.

Due to the numerical cost of the detailed size distribution
evolution procedure, $\langle e \rangle$ is assumed fixed, independent
of position and time. The first independency is justified as we
are considering a relatively narrow birth ring. The time independency 
is an acceptable simplification when considering that
the eccentricity dispersion is imposed by the largest
bodies present in the system \citep[e.g.][]{quillen07b} and that,
according to our current understanding of debris discs, i.e. systems
in which the bulk of the planetesimal accretion process is already over,
these bodies should be large planetary embryos
that are too big and isolated to be significantly affected by collisional
erosion over the $\sim 10^{7}-10^{8}$yrs timescale considered
here \citep[see, e.g.][]{wyat07,loeh07}. This is also why these
dynamics-imposing embryos are left out of the collisional cascade
numerically studied here \citep[for more on the subject, see][ as well
as the discussion in Sec.\ref{sec:thin}]{theb07}.

\begin{table}
\caption[]{Nominal case set up. The fields marked by a $^{*}$
are explored as free parameters in the simulations.}
\label{init}
\begin{tabular*}{\columnwidth} {ll}
\hline
Radial extent of birth ring* & $80<r<120\,$AU\\
Number and radial width of annuli* & 6$\times 6.66$AU\\
Initial surface density within ring & $uniform$\\
Total ``dust'' mass $M_{\rm dust}$ ($s<1\,$cm)$^{*}$& $0.1\,M_{\oplus}$\\
Stellar spectral type$^{*}$ & A5V\\
Radiation Blow-out size$^{*}$ & $s_{0.5}=5\,\mu$m\\
Size range modelled & 2$\mu$m$<s<$10\,km\\
Number of size bins & 99\\
Initial size distribution$^{*}$ & $dN \propto s^{-3.5}ds$\\
Dynamical excitation$^{*}$ & $\langle e\rangle_{0} =0.1=2\,\langle i\rangle_{0}$\\
\hline
\end{tabular*}
\end{table}

For the sake of clarity, we consider a nominal case (Tab.\ref{init}) with
set-up matching as closely as possible the $\beta$ Pictoris system,
i.e.  $M_{dust}=0.1M_{\oplus}$ and $\langle e \rangle=0.1$
\citep[e.g.][]{aug01}. For the radiation pressure blow out size, we take
$s_{0.5}=5\mu$m,the value derived for compact silicates
around a $\beta$ Pic like A5V star.
The system is divided into two radial zones:
\begin{itemize}
\item The parent body zone, or ``birth ring'',
located between $r_{\rm in}$ and $r_{\rm out}$ with the center at
$r_{\rm BR}$ and divided into 6 annuli,
where we follow the collisional evolution of the whole solid
body population, from a maximum size $s_{\rm max}=10\,$km
that sits in the planetesimal size range to a minimum size
$s_{\rm min} = 2\mu$m below the radiation blow-out limit.  The
initial size distribution in the entire range is assumed to follow the
idealized collisional ``equilibrium'' distribution,
$dN(s)_0\propto s^{-3.5}ds$ \citep{dohn69}. Our runs show that this
choice is not crucial: in the relevant dust--size range, the size
distribution is quickly relaxed toward a new steady state with a
profile that is independent of the initial choice and that deviates
significantly from a Dohnanyi-like power law (see section
\ref{sec:anal} and Fig.\ref{totcomp}).  
\item The outer zone, which is devoid of particles at the beginning
of the runs and gets progressively populated by small grains
coming from the birth ring. Consequently, we only follow here grains
with $s \leq s_{\rm max}^{*}$, where $s_{\rm max}^{*}$ roughly
corresponds to the biggest grains able to leave the parent body region
and is taken conservatively to have a radiation--to--gravity--ratio of
$\beta_{(s_{max}^{*})}=0.1$. Spatially, this region is divided into 3
annuli just outside the main ring plus one additional,
infinitely extended ``buffer'' annulus. Within the latter zone, no
collisional evolution is modeled, and only the orbital evolution of
the grains is considered: either escape of the system for unbound
grains or progression to the apoastron and return to the inner annuli
for the bound ones.
The radial extent of the outer zone (not including the ``buffer'' annulus)
is set to $\sim 4 r_{BR}$, which is typically the extent
of the``outer'' region considered for the 2 most famous
birth-ring/outer-zone systems, i.e., $\beta$ Pic and AU Mic
\citep[e.g.][]{aug01,stru06}.
\end{itemize}

The radial surface brightness profile in scattered light is then
synthetically computed using the dust size and radial distributions,
assuming an $r^{-2}$ dilution of the stellar flux and isotropic
scattering (although different scattering properties are also
explored).

We explore around the nominal set-up (see Tab.\ref{init}),
especially for the two fundamental parameters which are
$M_{dust}$ and the dynamical excitation of the system (as
parameterized by $\langle e \rangle$).

\section{Numerical Results}

\subsection{nominal case: smooth edge and $r^{-3.5}$ profile}

\begin{figure}
\includegraphics[angle=0,origin=br,width=\columnwidth]{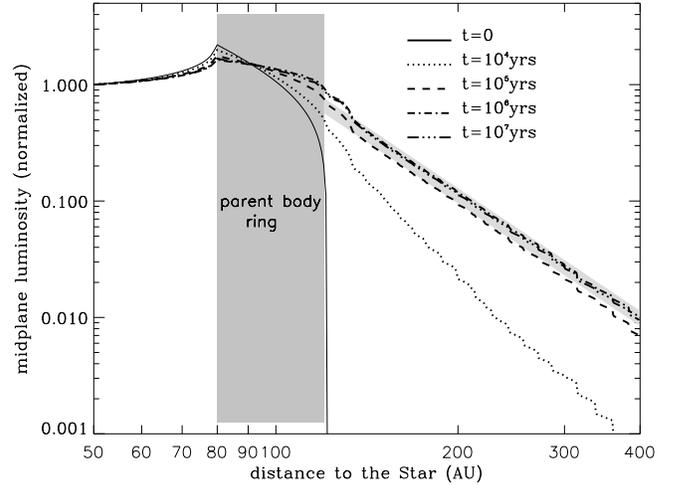}
\caption[]{Evolution of the midplane luminosity profile in scattered light.
Nominal case: parent body ring $[80,120]$AU, $M_{\rm dust}=0.1M_{\oplus}$,
$\langle e \rangle=0.1$. Each midplane luminosity has been
renormalized to its value at 50\,AU.  The dark grey area represents
the parent body region where all mass is initially located. 
The narrow light grey area represents a $r^{-3.5}$ slope reconstructed
backwards from the final luminosity value at 400AU with a $\pm 15\%$
width.}
\label{midnom}
\end{figure}
\begin{figure}
\includegraphics[angle=0,origin=br,width=\columnwidth]{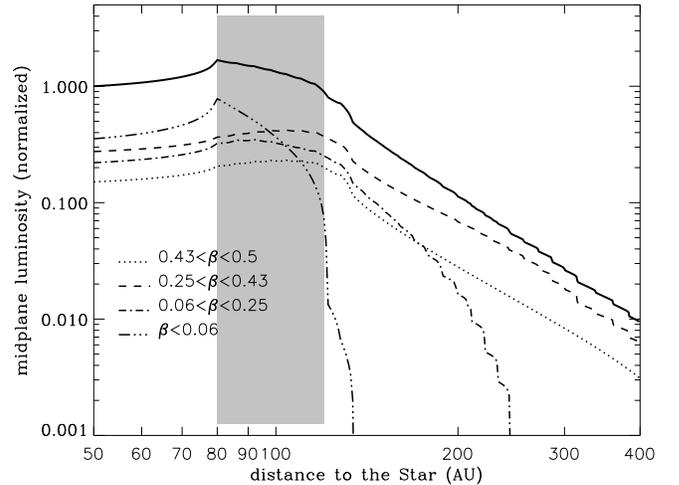}
\caption[]{Nominal case at $t=10^{7}$yrs.
Respective contributions of different grain populations (parameterized
by their $\beta$ value) to the total scattered flux, as functions of
the radial distance.}
\label{betnom}
\end{figure}
\begin{figure}
\includegraphics[angle=0,origin=br,width=\columnwidth]{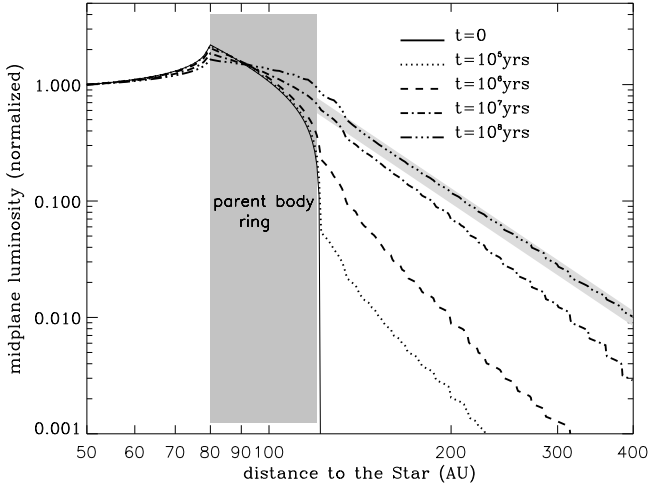}
\caption[]{Same as Fig.\ref{midnom}, but with 
$M_{\rm dust}=0.001M_{\oplus}$, all other parameters being the same.
Note that the final timescale is here $t=10^{8}$yrs.  }
\label{midlow}
\end{figure}

Fig.\ref{midnom} presents results obtained for our $\beta$ Pic--like
nominal case. As can be clearly seen, the system rapidly ($\sim
10^{5}$yrs) reaches a steady state after which the SB profile no
longer evolves significantly. We stop the integration at
$10^{7}$yrs.
This $\sim 10^{5}$yrs timescale is the time it takes for the small grains 
that fill the outer radial zone to reach an equilibrium between
collisional production and destruction.
A direct consequence of this fast evolution is that the
initial sharp outer edge of the parent body region is quickly smoothed
out. Once the steady state is reached, the profile in the
$r>r_{\rm out}$ region lies very close to $SB(r)\propto r^{-3.5}$.

When looking at the respective contributions from different dust
populations to the total profile, it appears that the scattered flux
is, in the 120--400AU region, dominated by high--$\beta$ grains in the
$0.25<\beta<0.4$ range (Fig.\ref{betnom}). These grains have orbital
eccentricities in the 0.33--0.75 range and apoastron in the 240--700AU
region and thus spend a large fraction of their orbits in the domain
located between $r_{\rm out}$ and $4r_{\rm BR} $.  Grains with even
higher $\beta$ (close to 0.5) only weakly contribute to the
flux in the $r_{\rm out}$ to $4r_{\rm BR} $ region because they have
orbits whose apoastron is often at several 1000AU and will
thus spend most of their time outside the $\leq 400$AU region
considered here.

For a case with a much less massive ($M_{dust}=0.001M_{\oplus}$) disc,
we see that the
system reaches a similar outer SB profile (SB $\propto r^{-3.5}$
outside $r_{\rm out}$) but with a much longer timescale: a few $
10^{7}\,$yrs instead of $\sim 10^{5}\,$yrs in the nominal run
(Fig. \ref{midlow}).
Similar SB profile is also obtained for a higher mass run ($M_{\rm
dust}=1M_{\oplus}$). Once again, the only difference with the nominal
case is the pace at which the steady--state is reached (almost
$10^{4}$yrs here).

\begin{table}
\begin{minipage}{\columnwidth}
\caption[]{Results obtained for different values of the system's
dynamical excitation $\langle e \rangle$
(\,=$\,2\,\,\langle i \rangle\,$)
and width of the birth ring $\Delta r_{\rm BR}$ \footnote{when varying its
outer edge $r_{\rm out}$, its inner edge being fixed at 80AU}. Also
shown are results for the nominal run but with the SB computed using
a different scattering phase function (see text for details)}
\label{results}
\renewcommand{\footnoterule}{}
\begin{tabular*}{\columnwidth} {lcc}
Run &   Outer edge sharpness \footnote{luminosity drop at the outer edge of
the ring, as measured by the flux ratio between the outer edge of the
birth-ring and 10AU beyond it} & \emph{SB} profile \footnote{average value in the
$1.5r_{BR}-4r_{BR}$ region}\\
 &   $SB(r_{\rm out})$/$SB(r_{\rm out+10})$ & slope $\alpha$\\
\hline
nominal case & 2.50 & -3.51 \\
$<e>=0.2$ & 2.44 & -3.48\\
$<e>=0.035$ & 2.78 & -3.52\\
$<e>=0.01$ & 3.33 & -3.49\\
$\Delta r_{\rm BR}
=20$AU & 2.05 & -3.51 \\
$\Delta r_{\rm BR}
=10$AU & 1.96 & -3.54 \\
\hline
scattering anisotropy g=0.5 & 2.22 & -3.68 \\
scattering anisotropy g=0.8 & 2.22 & -3.98 \\
\hline
\label{results}
\end{tabular*}
\end{minipage}
\end{table}

We further explore the parameter dependences by running a series of
simulations varying the width of the birth ring and the dynamical
excitation of the system. We also investigate the importance of the
scattering function assumption by exploring anisotropic cases,
assuming a \citet{hen41} phase function and changing the asymmetry factor
$\mid g \mid$. The results for all these runs are summarized in
Tab.\ref{results}, showing the 2 main outcomes of interest for the
present problem: the luminosity drop at the outer edge of the ring and
the average slope of the luminosity profile in the 1.5$r_{\rm BR}$--4$r_{\rm BR}$
region.  The nominal results are robust: the luminosity drop at the
ring's outer edge is always comprised between 2 and 3, whereas for the
slope we get $\alpha = -3.51\pm 0.03$.  The only way to reach a
somehow steeper slope is to assume a strong anisotropy of the
scattering function. But even for a rather extreme g=0.8 case, $\alpha
\sim -3.98$.

\subsection{``extreme cases'' producing different profiles} \label{sec:extreme}

\subsubsection{very massive disc} \label{sec:massive}

\begin{figure}
\includegraphics[angle=0,origin=br,width=\columnwidth]{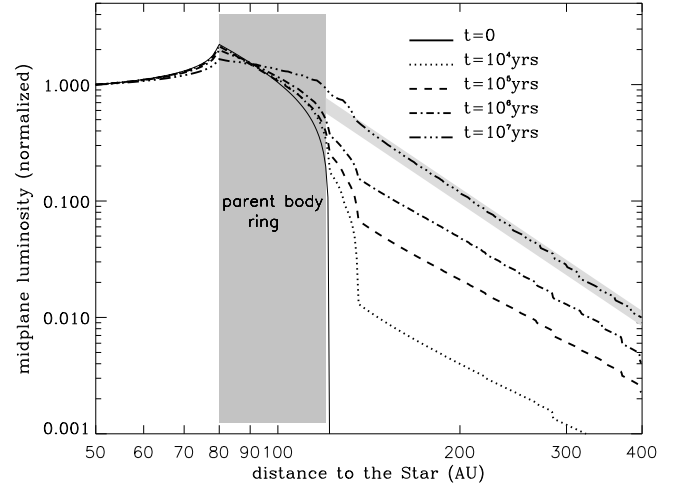}
\caption[]{Same as Fig.\ref{midnom}, but with a very high dust mass 
of $M_{\rm dust}=10M_{\oplus}$. }
\label{midvhigh}
\end{figure}

In order to push our parameter exploration to the limits, we keep
increasing the total initial dust mass $M_{\rm dust}$ until we observe
a departure from the nominal case SB profile. Such a departure is
observed for $M_{\rm dust}=10M_{\oplus}$ (Fig.\ref{midvhigh}). This
extremely massive case distinguishes itself from the others by the
fact that a sharp luminosity drop is maintained at the outer edge for
$\sim 10^{6}$yrs.   
What keeps the outer regions at a very low luminosity is the fact that
the parent body ring is so densely populated that grains pushed
outward by radiation pressure cannot freely escape it without
experiencing a collision.  This can be illustrated by looking at 
a simplified parameter, i.e. the geometrical
radial optical depth, defined (at a given distance from the star
$r_0$) by
\begin{equation}
\tau_{\rm rad}(r_0)=
\sum_{ia=i0}^{n_a} 
\frac{\int dN_{ia}/ds(r)
\pi s^{2}ds}{2\pi r H} \label{radial}
\end{equation}
where $dN_{ia}/ds(r)$ is the differential number of $s$--sized particles in
a radial annulus of index $ia$ centered at radial distance $r$, $H$
is the vertical height of the disc at that distance and
$i0$ is the annulus index corresponding to the radial distance $r_0$.
$\tau_{\rm rad}$ is of course simply the optical thickness of the disc to
stellar photons. Since particle orbits are never straight radial
lines, this quantity is only a first approximation of
their real in-plane (or horizontal) collisional probability.
However, for the smallest grains
this is a relatively good first-order approximation (see the more thorough
discussion on horizontal and vertical collision probabilities
in Sec.\,\ref{subsec:escape}).

\begin{figure}
\includegraphics[angle=0,origin=br,width=\columnwidth]{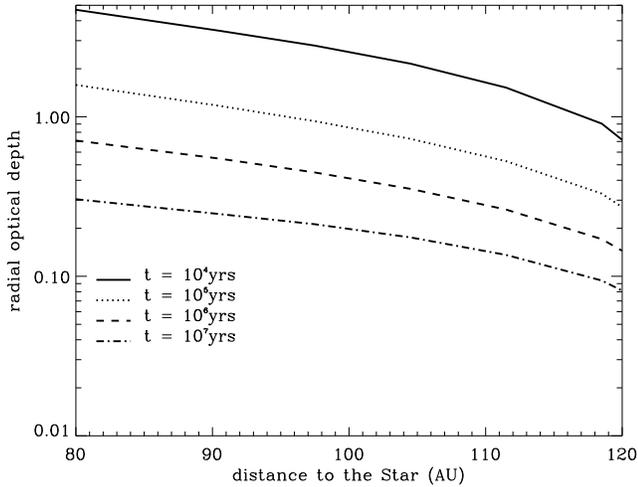}
\caption[]{Evolution with time of the radial optical depth (defined as in
eq. [\ref{radial}]) for the very high dust--mass case $M_{\rm
dust}=10M_{\oplus}$.}
\label{rad10M}
\end{figure}

Fig.\ref{rad10M} shows that $\tau_{\rm rad} \geq 1$ over most of the
source ring\footnote{When this is the case, one should also include
the attenuation of stellar light over the disc to be
self-consistent. However, we believe doing so would not qualitatively
alter our conclusion.} in the early epoch, so that few grains can escape
the birth ring without colliding with another grain, hence the density
(and luminosity) depletion in the $r>r_{\rm out}$ regions. However, this
radial optical depth steadily decreases over time, due to rapid mass
erosion by energetic collisions within the birth ring.
It drops below unity after $\sim 10^{5}\,$yrs,
and is $\sim 0.6$ by the time ($\sim 10^6$\,yrs)
the sharp outer edge is smoothed out (see Fig.\ref{midvhigh}).
After that, the system behaves like the nominal case.
We re-examine this radially optically thick
case in more detail in Sec.\ref{subsec:escape}.

\subsubsection{dynamically cold system} \label{sec:cold}

\begin{figure}
\includegraphics[angle=0,origin=br,width=\columnwidth]{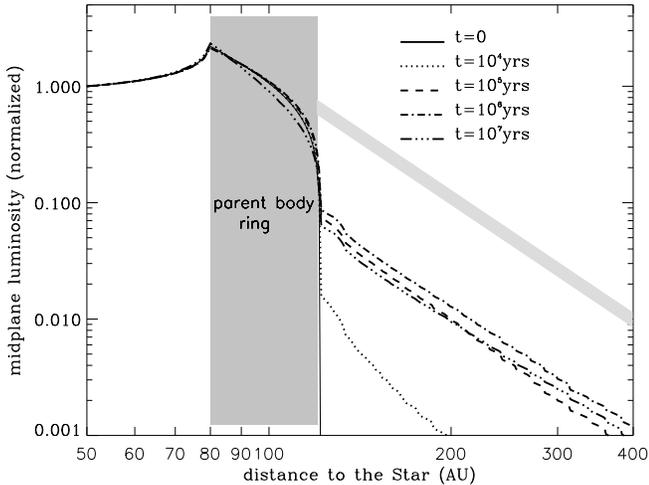}
\caption[]{Same as Fig.\ref{midnom}, but for a dynamically ``very cold''
system with $\langle e \rangle=0.001$.}
\label{midcold}
\end{figure}
\begin{figure}
\includegraphics[angle=0,origin=br,width=\columnwidth]{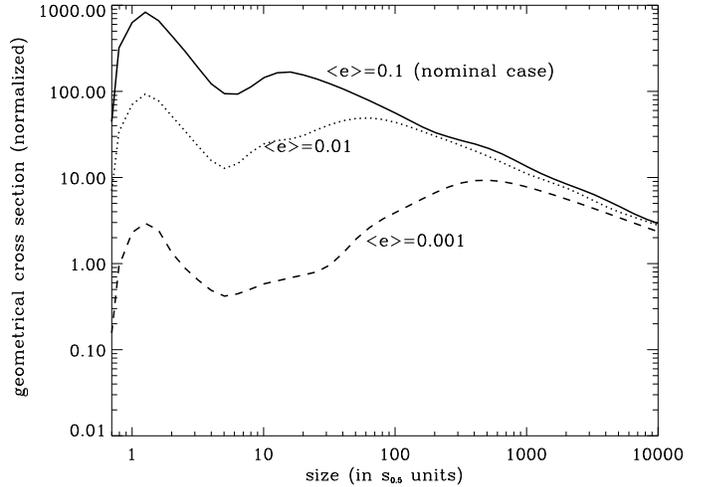}
\caption[]{Geometrical cross section per logarithmic size range,
integrated only for grains
inside the birth ring, for systems with different dynamical
excitation. These curves at obtained at the end of the integration
($10^7\,$ yrs) and are normalized to have the same value at the large
grain size.
}
\label{totopt}
\end{figure}

Another way of obtaining a departure from the standard SB
profile is to decrease the dynamical excitation of the system to a
very low value, typically $\langle e \rangle \leq 0.01$.
Fig.\ref{midcold} shows the results for the case
$\langle e \rangle=0.001$ ($\,=\,2\langle i \rangle$),
where a sharp outer edge is maintained
outside the birth ring for the entire duration of the simulation
($10^{7}\,$yrs).

The abrupt luminosity fall-off at the outer edge is here
due to a global depletion of small grains
directly resulting from an imbalance between the collisional
production and destruction rates of small and large particles in such a
dynamically cold disc.
This can be understood in the following way:

For a given dust mass, a lower dynamical excitation 
does not change the collision rate between large objects
not affected by radiation pressure. Indeed, because of
the equipartition $\langle e \rangle = 2 \langle i \rangle$
the decrease in $\langle \Delta v \rangle$ is exactly compensated by
the increase of the particle number density due to the reduced
thickness. However, the lower $\langle \Delta v \rangle$ values
mean that collisions will be less destructive and produce
less smaller fragments. This means that the rate at
which small grains (the ones significantly affected
by radiation pressure) are \emph{produced} is significantly
reduced.
On the contrary, the rate at which these small grains are $destroyed$
is higher. Indeed, collisions velocities for impacts involving
them are not significantly reduced by the small $\langle e \rangle$
for parent bodies, since small grain dynamics is predominantly 
imposed by radiation pressure. Furthermore, the rate at which
such impacts occur is increased, compared to the nominal case, because
of the increased radial optical depth (see Sec.\ref{sec:thin} for
a more detailed analysis).

As $\langle e \rangle$ decreases, this imbalance becomes more severe and
the depletion of the smallest grains is more acute.
This is illustrated in Fig.\ref{totopt} displaying, for
different values of the system's dynamical excitation,
the respective contributions of
different grain sizes to the total geometrical cross section $\sigma$.
For the nominal case ($\langle e \rangle=0.1$), we obtain the standard
result that $\sigma$ is dominated by the smallest grains close to the
cut--off size. As $\langle e \rangle$ gets smaller, however, the
contribution of these smaller grains progressively decreases. 
Below the limiting value $\langle e \rangle \sim 0.01$,
this effect is so pronounced
that the system's optical depth is no longer dominated by the smallest
grains, but by much bigger particles in the 100-1000$s_{0.5}$
range. These particles have their orbits largely confined within the
birth ring. This explains why a sharp luminosity decrease is observed
at the outer edge of the birth ring.

Note that contrary to the very high mass case, the sharp outer edge does not
smooth out with time but persist throughout the $10^{7}$yrs of the
simulation.

\section{Analytical derivation: The 'universal' $r^{-3.5}$ profile}
\label{sec:anal}

Our numerical exploration has shown that the $r^{-3.5}$ surface
brightness profile beyond the outer edge seems to be the most generic
outcome for a collisional ring system under the action of
stellar radiation pressure. We reinvestigate this issue from an
analytical point of view and derive simplified formulae confirming
this result.
We take here as a basis the analytical approach of \citet{stru06}
and extend it to more general cases regarding grain size and spatial
distributions.

We firstly assume that,
\emph{within the birth ring}, particles follow a power-law size distribution 
of index $q$ (instead of fixing $q=-3.5$ for a Dohnanyi equilibrium):
\begin{equation}
{dN_{\rm BR}} \propto s^{q} ds.
\label{eq:defineq}
\end{equation}
Since these grains spend most of their orbits in the empty region outside
the birth ring, their total number integrated
over {\emph the whole system}
($N_{\rm tot}$) will be boosted by a factor $1/f(e)$, where $f(e)$ is
the fraction of an orbital period a body of eccentricity $e\sim
\beta/(1-\beta)$ spends within the birth ring \citep{stru06}, so that
\begin{equation}
dN_{\rm tot} = \frac{1}{f(e)}dN_{\rm BR} \propto \frac{1}{f(e)}s^{q} ds.
\label{Eqstrub}
\end{equation}

We also follow \citet{stru06} in making the simplifying but reasonable
assumption that all high-$\beta$ grains are on average mostly seen
near their apoastron. Thus, at each given distance $r$ from the star,
the optical depth is dominated by particles of size
\begin{equation}
s_{dom}(r) = \frac{1}{1 -\frac{r_{\rm BR}}{r}}\,\,\,s_{0.5}.
\label{sdom}
\end{equation}
Since we are interested in the region $r \leq 4 r_{\rm BR}$, only
grains with their apoastron
\begin{equation}
a(1+e)=\frac{1-\beta}{1-2\beta} 
\left(1+\frac{\beta}{1-\beta}\right) r_{\rm BR}
=\frac{r_{\rm BR}
}{1-2\beta}< 4 r_{\rm BR}
\label{epos}
\end{equation}
are important. This corresponds to $\beta<0.4$ and thus $e<0.67$.
Note that, in their analytical derivation
(this assumption was relaxed in their Monte-Carlo model),
\citet{stru06} only considered
extremely eccentric $e\sim1$ grains, which spend most of
their orbits outside the considered $\leq 4 r_{\rm BR}$ region.

To estimate the enhancement ratio $1/f(e)$, we do not
adopt the asymptotic expression (valid for $e\lim 1$)
$f(e)\propto(1-e)^{1.5}$ taken
by \citet{stru06} but instead derive this value from Kepler's
equation for a typical particle produced at the middle of the birth
ring:
\begin{equation}
f(e) \propto (E_2 - E_1) - e(sin(E_2)-sin(E_1))
\label{kep}
\end{equation}
where 
\begin{equation}
E_1=0\\
{\rm and}\\
E_2 = acos\left[
\frac{1}{e}\,\left(1\,\,\,-\,\,\,\frac{r_{\rm BR}
+\Delta r_{\rm BR}
}{r_{\rm BR}
}\right)\right].
\label{E2}
\end{equation}
The vertical geometrical optical depth then scales with radius as
\begin{eqnarray}
\tau_{\bot}(r) & \propto & \frac{1}{r} \frac{dN_{tot}(s_{dom})}{dr} 
\,s_{dom}^{2}(r)
\nonumber \\
& \propto & \frac{1}{f(e)} \frac{1}{r} \frac{dN_{BR}(s_{dom})}{dr}
\,s_{dom}^{2}(r).
\label{opt}
\end{eqnarray}
Using the radial dependence of $s_{dom}$ given by eq.\ref{sdom}, we
obtain
\begin{equation}
\frac{dN_{\rm BR}(s_{dom})}{dr} \propto 
\left(1-\frac{r_{\rm BR}
}{r}\right)^{-q-2} \frac{r_{\rm BR}
}{r^{2}}.
\label{Ndep}
\end{equation}
Inserting Eqns.\ref{Eqstrub},  \ref{sdom} and \ref{Ndep} into
Eq.\ref{opt} leads to
\begin{equation}
\tau_{\bot}(r) \propto
\frac{1}{f(e)}\left(1-\frac{r_{\rm BR}
}{r}\right)^{-q-4} \frac{r_{\rm BR}
}{r^{3}}
\label{taufit}
\end{equation}
where $f(e)$ is given by solving Equ.\ref{kep}.

\begin{figure}
\includegraphics[angle=0,origin=br,width=\columnwidth]{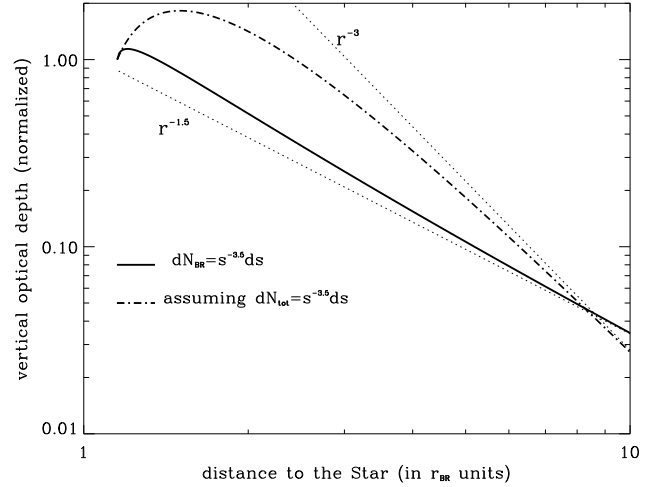}
\caption[]{Radial profile of the vertical optical depth,
obtained with our analytical fit of eq.\ref{taufit}, assuming 
$q=-3.5$ (with a ring width $\Delta
r_{\rm BR} =0.1r_{\rm BR}$). As a comparison, we also plot
the profile obtained with the (incorrect) assumption that the size
distribution is $dN_{\rm tot}\propto s^{-3.5}ds$ \emph{in the whole
system}.}
\label{fit}
\end{figure}

As a test, we first consider the same fiducial case as \citet{stru06}, i.e.
the Dohnanyi $q=-3.5$ value.
The radial profile of $\tau_{\bot}(r)$ obtained this way
scales as $r^{-1.5}$ (Fig. \ref{fit}). Departures from
this slope are relatively limited, even in the 1-1.5$\,r_{\rm BR}$
region, where $\tau_{\bot}(r)$ has a $\sim r^{-1.75}$ dependence.  If
we apply the usual rule of thumb that for an optical depth profile
$\tau_{\bot}(r) \propto r^{\alpha}$, the midplane SB scales as
$r^{\alpha-2}$\citep[e.g.][]{nak90}\footnote{This assumes grey
scattering and that the disc's vertical thickness scales as $H\propto
r$.}, we obtain $SB(r)\propto r^{-3.5}$ in most of the outer region,
confirming the result of \citet{stru06}.
For comparison, we
also plot on the graph the $\tau_{\bot}(r)$ profile derived when
(incorrectly) assuming that the Dohnanyi law applies to the entire
system, i.e., $dN_{tot}\propto s^{-3.5}ds$
\citep[as was done in][]{lec96,aug01,theb05}. Not surprisingly,
we recover the asymptotic dependence $\tau_{\bot}(r) \propto r^{-3}$,
corresponding to the $SB(r) \propto r^{-5}$ slope derived in these
past studies.

\begin{figure}
\includegraphics[angle=0,origin=br,width=\columnwidth]{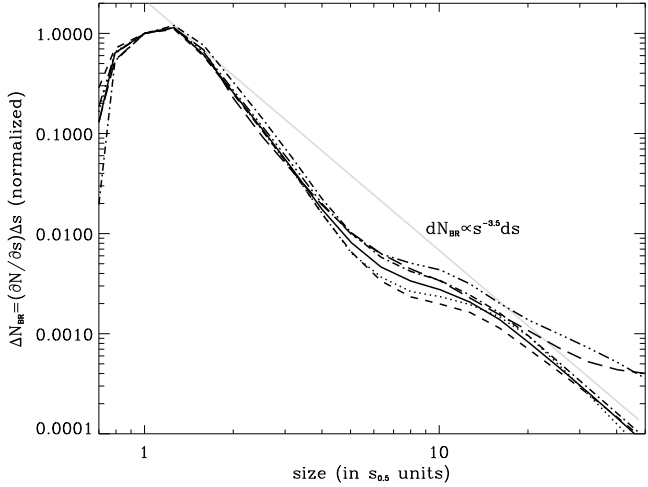}
\caption[]{Grain size distribution inside the birth ring ($dN_{\rm BR}/ds$), 
for all cases presented in Sec \ref{results}.
The size range here covers that of grains dominating the optical depth
in the outer region.  In order to facilitate comparison, all curves
have been renormalized to their peak value at $s=1.2s_{0.5}$.  The
thick grey line shows the theoretical Dohnanyi profile $dN_{\rm
BR}\propto s^{-3.5}ds$.  }
\label{totcomp}
\end{figure}

\begin{figure}
\includegraphics[angle=0,origin=br,width=\columnwidth]{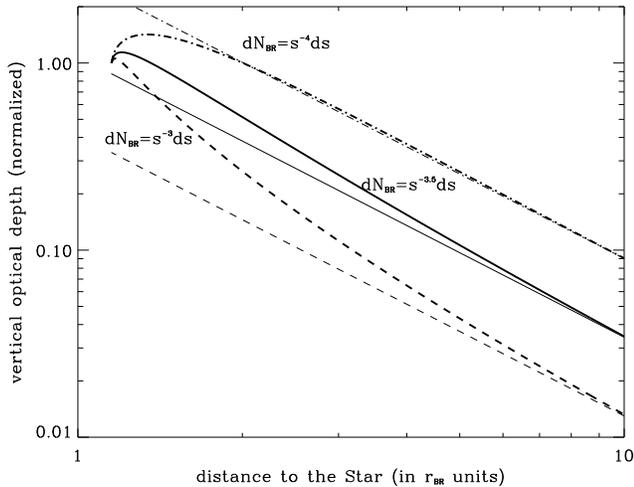}
\caption[]{Same as Fig.\ref{fit} but for different size distributions
index $q$. All size distributions lead to $\tau_{\bot} \propto
r^{-1.5}$ at large distances, but steeper size distributions
approaching this asymptote earlier.  }
\label{fitg}
\end{figure}

However, as has been often pointed out \citep[e.g.][]{theb03},
size distributions follow a power law
with index $q=-3.5$ only in unrealistically idealized systems,
with an infinite size range and no size dependence for
collisional processes. In our
simulations, the steady state size distribution within the birth ring
significantly departs from this value (see Fig.\ref{totcomp}).
This departure from the Dohnanyi law, especially in the
crucial size domain of the smallest grains, is a well know results
which has been obtained and discussed in several previous studies
\citep[e.g.][]{bag94,theb03,kriv06,theb07,loeh07}. The main reason
for this behaviour is that radiation pressure
introduces a natural minimum cut-off $s_{0.5}$ in the size
distribution, so that bodies of size $s_1$ just above $s_{0.5}$ are
overabundant because of the lack of small $s<s_{0.5}$ impactors that
could destroy them. Subsequently, larger bodies, of size $s_2$, which
can be destroyed by impacts with $s_1$ ones, are underabundant, which
in turn leads to an overabundance of objects of size $s_3>s_2$, {\it
etc}. The resulting size distribution displays a pronounced
wavy-pattern observable also in Fig.\ref{totcomp}.
For the size range which is here of special interest, i.e. grains
in the $0.15<\beta<0.4$ ($1.25s_{0.5}<s<3.5s_{0.5}$) range,  
the distribution is steeper than a Dohnanyi one and is close to $q\sim
-4$. Interestingly, putting this value into Eq.\ref{taufit}
leads to a $\tau_{\bot}(r)$ profile
that is again well (in fact, better) approximated by the $r^{-1.5}$ slope
(Fig. \ref{fitg}). This is because, in eq.\ref{taufit}, the $r$
dependence arising from the size distribution term $(1-r_{\rm BR}
/r)^{-q-4}$ is swarmed by that from the geometric term ($r_{\rm BR}
/(f(e)\times r^{3})$ at large $r$.  Only the rapidity at which the
$r^{-1.5}$ asymptotic behaviour is reached depends on the index $q$
(Fig. \ref{fitg}).

\section{Discussion} \label{sec:discussion}

\subsection{How to escape the universal $r^{-3.5}$ profile}
\label{subsec:escape}

\begin{figure}
\includegraphics[angle=0,origin=br,width=\columnwidth]{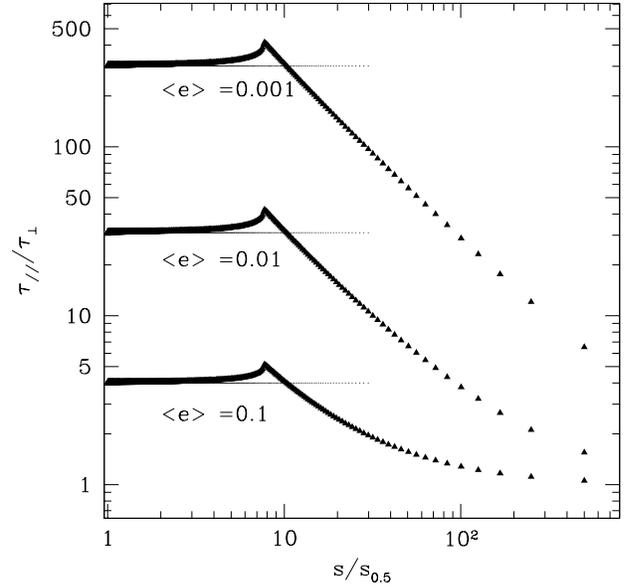}     
\caption{Ratios between the horizontal and vertical 
optical depth plotted as a function of grain size for
different values of dynamical excitation.
The straight lines associated with each case mark the values
of $\tau_{\rm rad}/\tau_{\perp}$ (see eq. \ref{radial}). The grains
are assumed to be launched with local Keplerian circular velocity at
the centre of the ring.  The fractional ring width $\Delta r_{\rm
BR}/r_{\rm BR}$ takes the nominal value of $0.3$. Decreasing this
value will move the peaks towards the right.}
\label{fig:taue-HR} 
\end{figure}

Our numerical explorations show that there are only 2 ways \emph{not} to
end up with the standard result with no sharp edge
and $SB \propto r^{-3.5}$: a very massive, radially optically thick
disc or a dynamically cold system.
For these 2 ``extreme'' cases, the analytical derivation of the
previous section becomes invalid. Although the reasons why the
analytical study no longer holds are different for each case, they
are nevertheless for both cases distinct consequences of the same
crucial characteristics of the system's dynamics when
pushed to its limits: the strong imbalance between the collision rates
and velocities of large and small particles. 

Therefore, before discussing these 2 cases in more detail, let us
present the mechanisms at play behind this 
large/small particle dichotomy.

For large bodies, we have seen that collision velocities can be 
derived using the standard expression for their dynamical rms excitation
(Eq.1). Since for these objects
equipartition between in-plane and off-plane motions results
in a fixed $\langle e \rangle /  \langle i \rangle$ ratio,
it follows that collision \emph{velocities} are directly proportional
to the system's inclination whereas
collision \emph{rates} are independent of $\langle i \rangle$.
That is to say that
the collisional optical depth is directly proportional to the
vertical optical depth $\tau_{\perp}$ (which does not vary with
$\langle i \rangle$).
This is no longer the case for small grains pushed on eccentric
orbits by radiation pressure, which can sample
the whole radial extent of the birth ring.
Radiation pressure induced eccentricities are equal to
\begin{equation}
e_{\beta} = \left(1-
\frac{(1-2a \beta/r)(1-\langle e^2  \rangle)}{(1-\beta)^2}
\right)^{1/2},
\label{eq:ebeta}
\end{equation}
where $a$ is the semi-major axis of the parent body producing
the small grain, and $r$ the radial location of its release.
For these objects, in-plane
motions become much more important than vertical ones. This can be
quantitatively estimated by computing the in-plane collisional
optical depth $\tau_{//}$. This quantity is obtained by
integrating, for a given particle, over distances that are
travelled by this grain relative to its background,
\begin{equation}
\tau_{//} = \oint {|{\bf v} - {\bf v_0}|}dt\, n \sigma,
\label{eq:deftau2}
\end{equation}
where ${\bf v}$ is the grain velocity, ${\bf v_0}$ the mean velocity
of particles in the region, which we approximate to the local
circular Keplerian velocity.  $n$ and $\sigma$ are the local number
density and grain cross section, and their product is related to the
vertical optical depth by $\tau_{\perp} = 2 H n\sigma$. The
integration ($\oint$) is over the whole orbit and $n\sigma$ is
non--zero only inside the birth ring. 
For a given particle $\tau_{//}$ depends on 2 main parameters:
its orbital eccentricity and, for the smallest grains,
the birth ring's radial extent (since this
is the only region where collisions can occur).
For large objects, equipartition between
$\langle e \rangle$ and $\langle i \rangle$ means that
$\tau_{//}$ is simply equal to $\tau_{\perp}$ (see Fig.\ref{fig:taue-HR}),
which is another way to say that impacts in the in-plane and
vertical directions are equally important.
As grains get smaller, however, the ratio $\tau_{//} /\tau_{\perp}$
increases, reaching a maximum when the grain's orbit has the
same radial extent as the birth ring, hence the peak at
$s \sim 7\,s_{0.5}$ for the nominal birth ring
considered in Fig.\ref{fig:taue-HR}.

The imbalance between small and large grains collision rates
is directly proportional to $e_{\beta}/\langle e \rangle$.
Since $e_{\beta}$ only weakly depends on $\langle e \rangle$
(Eq.\ref{eq:ebeta}), this means that the imbalance increases
for dynamically cold systems, as clearly appears in Fig.\ref{fig:taue-HR}.

\subsubsection{radially optically thick disc}

In Sec.\ref{sec:massive} we identified one
consequence of the boosted collision rates of small particles, which is
that, for very high $M_{dust}$ values, high-$\beta$ particles have a
significant chance to experience a collision before escaping the birth
ring. In this situation, the analytical derivation of the
previous section is invalid because it implicitly assumes
that all high-$\beta$ grains produced in the birth ring
eventually reach their apoastron far in the outer regions.

To characterize this collisional optically thick case,
we used in Sec.\ref{sec:massive} the parameter $\tau_{\rm rad}$,
as defined in Eq.\ref{radial}, and empirically found that the limiting
$\tau_{\rm rad}$ value is around 0.6. 
The validity of the $\tau_{rad}$ parameterization is confirmed by
Fig.\ref{fig:taue-HR}, which shows that this simplified parameter
is actually a good approximation of the ``real'' in-plane (or horizontal)
collisional optical depth  $\tau_{//}$ for small grains close to the $s_{0.5}$
limit. We thus keep this easily defined parameter (as compared to $\tau_{//}$)
as a good approximate limiting criterion.
Writing as a first approximation that
\begin{equation}
\tau_{\rm rad}\sim \frac{\Delta r_{\rm BR}}{r_{\rm BR}}\, 
{1\over{\langle e \rangle}} \,
\langle \tau_{\perp} \rangle_{BR},
\label{radtau}
\end{equation}
we see that there are two ways of reaching high
$\tau_{rad}$ values: either by
increasing the total dust mass of the system (as parameterized by
$\tau_{\perp}$ or $M_{dust}$) or by decreasing its dynamical excitation.

We have seen in our numerical exploration that only playing on the
$M_{dust}$ (or $\tau_{\perp}$) parameter requires to reach values of
the order of $10 M_{\oplus}$ for our nominal
case of a disc with $\langle e \rangle =0.1$ and $\Delta r_{\rm
BR}/r_{\rm BR}=0.3$. However, our simulations show that global
collisional erosion in such a high-mass system is so intense
that the $\tau_{\rm rad}\geq 0.6$ regime can only be maintained for
$\sim 10^6$yrs before the birth ring becomes radially
transparent for high-$\beta$ grains, at which point the sharp outer edge
vanishes (see Fig.\ref{midvhigh}). 
One might argue that the largest
bodies in our simulation are in the 10\,km range and thus can be
quickly eroded in this high density case, while realistic systems
may have a large mass
reservoir contained in larger bodies and could sustain a longer period
of extreme dustiness. However, in this case the total
mass of solids in the system, which is already $M_{tot} \sim 10^4 M_{\oplus}$
in our high-mass run with $s_{max}=10$\,km, would reach unrealistically high values.
Another strong argument for ruling out this high-mass case is that
the presence of $10\,M_{\oplus}$ of dust is not supported by
observations. Indeed, no debris disc around a main sequence star seems to
contain more than $\sim 1M_{\oplus}$ of dust \citep{grea05}, or has a
vertical optical depth exceeding $\sim 10^{-2}$ \citep{kal05b}.

The other alternative for reaching high $\tau_{rad}$ values,
i.e. reducing the dynamical excitation, would
require, for a typical $M_{dust}=0.1M_{\oplus}$ disc, 
$\langle e \rangle$ values as low as $\sim 0.001$.
However, for such low values of the excitation, we have seen that
the system's collisional evolution and size distribution is already in 
the qualitatively different ``dynamically cold'' regime, where
small particles cease to dominate the optical depth (see next section).

\subsubsection{dynamically cold disc: depletion of small grains} \label{sec:thin}

This dynamically cold mode, with typically $\langle e
\rangle < 0.01$, is in fact the second way to maintain a sharp outer edge
(see Sec.\ref{sec:cold}).
In this case, the imbalance between small and large grains'
collisional behaviour gets very pronounced (as it appears clearly
in Fig.\ref{fig:taue-HR}), up to a point where
it causes a depletion of small high-$\beta$ grains.
We have seen in Sec.\ref{sec:cold} that this is because
small grains are predominantly created by impacts involving
large objects, which get less efficient for low
$\langle e  \rangle$, but predominantly destroyed by
collisions involving themselves, whose efficiency
(imposed by radiation pressure-induced motions) only weakly
varies with $\langle e \rangle$ of the parent bodies and whose rate
strongly increases for decreasing $\langle e \rangle$.
As a consequence, contrary to the nominal case
the system's optical depth and luminosity are
dominated by larger $\sim 100 s_{0.5}$ grains that do not leave the
birth ring, hence the sharp, and long-lived drop at its outer edge
\footnote{The analytical derivation of Sec.\,\ref{sec:anal}
is invalid here, since it implicitly 
assumes that small high-$\beta$ grains dominate the total optical
depth}.

However, we are here confronted with two issues. 

The first one regards the total brightness of such discs.
Indeed, a disc deprived from its smallest grain population
(in the $\leq 100s_{0.5} \sim 0.5$mm range) should appear much dimmer
than a disc in which these grains dominate the light scattering area.
From Fig.\ref{totopt}, we see that, for the same total mass of dust
(always contained in the biggest grains),
a small-grain-poor system with $\langle e \rangle \leq 0.01$
is at least 10 times less luminous
than the nominal case. It follows that, to reach the same fractional
luminosity (assuming it is proportional to the optical depth), the
dynamically cold system has to be at least 10 times more massive.
This has no direct implications on observed dust mass estimates,
which are usually derived from sub-mm or millimetre observations in thermal
emission, but should affect the correlation between $M_{dust}$ and total
fractional luminosities $f_L$. In this respect, dynamically colder
(or thinner) discs should have higher $M_{dust}/f_L$ ratios.
However, the uncertainties affecting both parameters' estimates are
probably high enough to accommodate a factor 10 uncertainty in their ratio
\citep[see for example the attempt at connecting these two quantifies
for the specific case of AU Mic performed by][]{aug06}, so that
our result is here probably not very constraining.

The second and probably crucial issue is how likely it is to
find such dynamically cold discs, with $\langle e \rangle$ and
$\langle i \rangle$ lower than $\simeq 0.01$.
As noted by \citet{theb07}: ``the only observational constraint
[on the disc's dynamical excitation]
comes from measuring the disc's vertical thickness and
deriving estimates of orbital inclinations, but such constraints
are scarce''. Edge-on discs represent the most favourable cases
since H/r can be directly measured. However, even for
the two most studied discs,
only partial information is available. Krist et al. (2005) find
$H/r \leq 0.04$ for AUMic (and $H/r \leq 0.02$
close to the position of maximum surface density). The
$\beta$ Pictoris disc appears thicker with H/r ratios as
large as $\simeq 0.1$ (Golimowski et al. 2006).
\footnote{However, these measurements
include the so-called disc warp which, according to
Golimowski et al. (2006), might be due to a blend of two separate,
intrinsically thinner disc components inclined with respect
to each other by a few degrees}.
The modelling and inversion of scattered light brightness profiles
of inclined, ring-shaped discs do not provide many more
constraints. The HD181327 rings for example,
might have H/r ratio as large as about 0.1 at the positions of
maximum surface density, but the actual ratios could be two
times smaller (Schneider et al. 2006).

There are thus large uncertainties, but it seems however that the
0.01 to 0.1 range is the most realistic one for
$\langle i \rangle$ (and thus $\langle e \rangle$ if assuming
equipartition).
This 0.01-0.1 range does also make sense when considering simple theoretical
arguments regarding the sizes of the biggest objects within the disc.
Consider indeed a belt of planetesimals sitting at $\sim 100$ AU from the
central star. If the maximum size is
limited to, say, $\sim 10$km, then excitation by mutual 
viscous stirring leads to values of the order of their surface escape
velocity ($\sim 10$m/s), and collisions among them generate particles
with dispersion velocity that does not exceed the same velocity.
The resulting $H/R
\sim v_{\rm esc}/v_{\rm kep} \sim 0.003$ and is dynamically cold. On
the other hand, if the maximum size is $\sim 500 $km, then
$H/R \sim 0.15$ and the resulting debris disc is dynamically
hot. This dynamically hot case probably makes more sense within the
frame of the standard planet formation scenario, in which debris discs
correspond to systems in which the bulk of planetesimal accretion process
is already over and large planetary embryos are present
\citep[e.g.][]{ken05}.

\subsection{Application to real systems} \label{sec:real}

As a consequence, we expect our nominal result, valid in the
$\langle e \rangle \geq 0.01$ range, to correspond to the
``natural'' collisional evolution of most discs when left to themselves.
This nominal model, with no sharp edge and luminosity falling as
$r^{-3.5}$ compares well against the outer region SB profiles
(beyond $\simeq 40$AU and $\simeq
120$AU respectively, see Table \ref{table:observed})
for the two perhaps most emblematic debris discs: $\beta$ Pic and AU Mic. 
For these two systems, which are amongst the few
ones for which (partial) disc thickness estimates are available, this result is in
good agreement with the observed $\langle i \rangle$, which is
for both discs in the $\geq 0.01$ range, and thus 
in the dynamically "hot" regime displayed in Fig.\ref{midnom}.\footnote{
Our fit cannot give more quantitative information on the
$\langle e \rangle$ and $\langle i \rangle$ dispersions, except
that they are above the $\sim 0.01$ threshold value for the dynamically cold mode.}
For AU Mic, this conclusion confirms the initial results obtained by
\citet{stru06} with a more simplified approach.
For $\beta$ Pic, this result should be reinvestigated in more detail in
a more complete study taking into account the numerous additional features
observed in this system, but we believe our result to be robust regarding
the global shape of this system's outer profile.
Within the observational error bars, the nonimal result characteristics are also
in relatively good agreement with outer disc profiles derived for
several other debris discs, i.e. HD53143 or HD32297 (Tab.\ref{table:observed}),
especially when taking into account that strong anisotropic 
scattering could lead to somehow steeper profiles in $r^{-4}$
(Tab.\ref{results}).

For all these systems, no additional sculpting mechanism is needed in
order to explain the main ring's edge and the profile beyond it.
Of course, something had to sculpt the \emph{parent-body} ring in the
first place, but such a sculpting could have occurred in the past,
when for instance gas was still present, and no longer be active today.
Another alternative is that the birth-ring's outer edge corresponds to
the natural radial limit at which large planetesimals can form in the
protoplanetary nebulae.

There is however a group of systems (see Table \ref{table:observed})
which have $SB$ slopes in the $-4.5$ to $-5$ range
(HD139664, HD107146, HD181327), or
even very sharp ring-like systems with brutal luminosity
decreases in $r^{-6}$ or $r^{-7}$, represented by Fomalhaut and HR4796,
which strongly depart from our nominal profile.
Does this mean that these system's outer edges cannot be produced
by ``natural'' internal collisional evolution and that ``something else''
is acting as a sculpting agent? 
The main question is here to see if these systems could fall into
one of the categories giving alternative profiles presented in
Sec.\ref{sec:extreme}, especially the self-sustaining dynamically cold case.
As a representative example, we consider the specific case of HR4796A.
\footnote{The imaging data
on Fomalhaut do not cover regions beyond $\sim 20$ AU from the outer
edge \citep{kal05}, which is not enough to proceed to a reliable
numerical fit}

\subsection{fitting a sharp edge ring: HR4796A}

\begin{figure}
\includegraphics[angle=0,origin=br,width=\columnwidth]{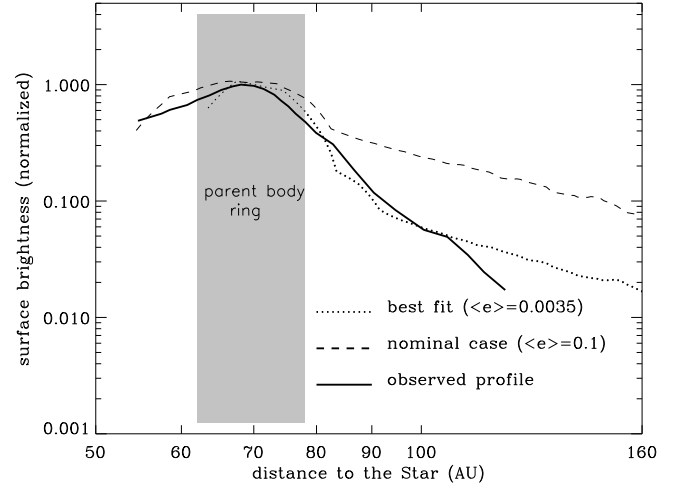}
\caption[]{The radial deprojected surface brightness profile
for the HR4796 disc.  Observed data are taken by \citet{schnei99} and
are displayed in Fig.6 of \citet{wahhaj05}. Numerical profiles have
been degraded to the resolution of the NICMOS/HST images, i.e.,
0.12''=8AU.  The best numerical fit is obtained for $\Delta r_{\rm BR}
$=16AU and $\langle e \rangle = 0.0035$.  }
\label{hr4796}
\end{figure}
As for most debris discs, there exists no fully reliable observational
constraints on the vertical height and therefore dynamical excitation
of the HR 4796A disc.  From model fitting of scattered light and IR
images, \citet{aug99} argue that the scale height at 70AU (the location
of the main ring) has to be less than 7-8AU, or $2\langle
i \rangle \sim \langle e \rangle \leq$ 0.1. 
However, from image fitting and theoretical constraints,
\citet{ken99} find that the most likely value is $H \sim 0.5$AU, i.e.
$2\langle i \rangle \sim \langle e \rangle \leq 0.007$. It is thus
plausible that this system falls into the dynamically cold category.

We numerically explore a series of dynamically cold HR4796-like
systems, fixing the ring's center at 70AU and taking the radial extent
of the birth ring $\Delta r_{\rm BR} $ as well as $\langle e \rangle$
as free parameters. The observed profile comes from HST/NICMOS
\citep{schnei99} and is displayed in Fig.6 of \citet{wahhaj05}.
Fig.\ref{hr4796} shows the best fit obtained in our parameter
exploration. It corresponds to a dynamically cold case with $\langle e
\rangle=0.0035$ and a radial width of the birth ring $\Delta r_{\rm
BR} $=16AU, which is close to the $\simeq 13-17$AU derived from
observations \citep{schnei99,schnei01} \footnote{Note that since the disc's
width is resolved in the near-IR \citep{schnei99,schnei01}, it could in
principle be taken as a fixed input in our model. However, we kept it as a
free parameter to check the validity of our fit.}.
As a comparison, we also display a
``nominal'' profile obtained for $\langle e \rangle=0.1$. The latter
is fully incompatible with the observed profile over the 80-120AU
domain by almost 2 magnitudes in brightness. The best fit (dynamically
cold case) compares well with the data in the 70-110AU region and
departs from the observed $SB(r)$ profile only between 110 and 120AU, where the
numerical profile becomes too flat. It is difficult to say how
significant this discrepancy in these outermost 10AU is, as there are
no error bars given in the \citet{wahhaj05} plot. It is possible that
confusion from the sky background enters at these distances.

We must thus remain careful but it seems that there is
at least a possibility
for our dynamically cold model to provide an explanation
for sharp outer edge discs like HR4796A.
The question of how realistic this explanation can be is
another issue. We have seen in Sec.\ref{sec:thin} that the
$\langle i \rangle<0.01$ condition, although probably not generic with respect
to planet-formation scenarios, cannot, in most cases, be explicitly
ruled out by observations.

There may however be a prediction made by the cold-disc scenario
which could be observationally checked, i.e., the underabundance
of grains in the $\mu$m to sub-millimetre range
\footnote{While this paper was being reviewed, \citet{liseau08}
reported the detection of a dust ring around q Eri, which
could possibly be very confined and depleted from $\leq 100\mu$m grains}.

For HR4796, for instance, \citet{aug99} have performed detailed fits
of the SED as well as of thermal and scattered light images. Their
best fit implied that most of the geometrical cross section was contained
in grains close to the minimum value for their size distribution, i.e.
$s_{min}\sim 10\,\mu$m. However, this fit was obtained \emph{assuming} an
imposed size distribution in $s^{-3.5}$, so that these results
cannot be used to rule out the possibility of higher $s_{min}$ values
for alternate size distributions.
\citet{wahhaj05} performed similar fits with partly more recent data
and found that the effective size for grains within the ring is $\sim 50\,\mu$m.
This would be in relatively good agreement with our dynamically cold case.
However, this value was obtained
assuming a single size for the dust population, an obvious limitation
the authors are fully aware of as they acknowledge the need for
``physically meaningful size distributions''.
It is a general characteristics
of such global fits to have too many free parameters to constrain
the dust composition as well as size and spatial distributions in an
unambiguous nondegenerate way.
They usually have to rely on some starting assumption regarding the size
distribution. In this respect,
it would be interesting to test these best fit models with
alternate size distributions, as obtained from numerical collision-evolution
studies, but this is an issue which exceeds by far the scope of the
present work.

There exists however yet unpublished scattered light data
(Schneider et al.,
{http://nicmosis.as.arizona.edu:8000}) indicating
that the ring is uniformly red from V to H. This should mean
that the emission is dominated by grains in the sub-$\mu$m to $\mu$m range.
Should this feature be confirmed, then the dynamically cold state could probably
be ruled out. 
Note however that this puzzling feature would be very difficult to interpret for
\emph{any} dynamical model of the HR4796A ring. Indeed, it is
not easy to explain how the luminosity, and thus
the geometrical cross section could be dominated by \emph{unbound} grains
which should leave the ring on dynamical timescales \footnote{This seems
to be a recurrent issue which is not limited to HR4796A: vast populations
of unbound particles could also dominate the scattered luminosity of
HD141569A \citep{aug04}, and might be present in the outer \bp\ disc
\citep{aug01}}.

\section{Summary and Conclusion}

We numerically investigate if the diverse outer surface brightness
profiles observed in debris discs can be explained
by the ``natural'' collisional evolution of belts made of solid particles
ranging from planetesimal to micron-sized dust grains.
We consider an initial ring of parent bodies with a razor-sharp
outer edge and quantitatively examine to what extent the steady collisional
production of small, radiation-pressure-affected grains modifies
the initially perfect ring-like structure.

Concurring with the pioneering results of \citet{stru06}, our
numerical explorations have shown that for most ``reasonable''
parameters (mass, ring width, dynamical excitation) of a collisionally
evolving debris ring, the surface brightness outside the ring
naturally tends towards a standard profile, with no sharp drop at
the outer edge and a mid-plane surface brightness profile $\propto r^{-3.5}$.
We confirm this result by simple but
robust analytical considerations.
This nominal result is in good agreement with the luminosity
profiles observed in the outer regions of several debris discs, including
\bp\ and AUMic.

Some observed systems however exhibit steeper radial luminosity profiles.
Our numerical exploration shows that sharper outer edges and
profiles steeper than that of the nominal case
can only be obtained for
two ``extreme'' cases:
\begin{itemize}
\item
1) For a system with very high dust content (typically
$M_{dust}\geq10M_{\oplus}$), the radial optical depth is raised to
near unity for small grains.
Most of these high-$\beta$ grains cannot travel out
of the birth ring without suffering a collision. This allows the outer
edge to sharpen. However, systems with such high radial optical depth
are expected to wear down with time because of strong collisional erosion.
Moreover, the existence of such very dusty systems is not
backed by observations. This case does not appear as a realistic
option.
\item
2) A longer survival of the sharp outer edge is achieved for
systems with normal dust content that are \emph{dynamically cold}, with typically 
$\langle e \rangle =2\langle i \rangle \leq 0.01$.
In this case, small grains are destroyed much more efficiently
than they are created, leading to a depletion of this population.
The system's optical depth and luminosity are then dominated by
large grains which do not leave the main birth ring, leading
to a sharp outer edge. Even if this case might not
correspond to the most generic debris disc configuration, it cannot
be ruled out by observations and is thus a possible explanation to
some of the observed systems.
\end{itemize}

To numerically investigate the applicability of the dynamically cold case
to real sharp-edge systems, we consider
the specific case of HR4796A. We find a reasonably good
fit of this system's outer region luminosity profile with
a dynamical excitation $\langle e \rangle \sim 0.0035$.
There is thus the possibility that such a sharp outer edge could be
explained by the natural collisional evolution of a confined disc
of large parent bodies.

\begin{acknowledgements}
PT wishes to thank Jean-Charles Augereau and Alexander Krivov
for fruitful discussions.
\end{acknowledgements}

{}
\clearpage

\end{document}